\documentclass[twocolumn,aps,british]{revtex4}
\usepackage[T1]{fontenc}
\usepackage[dvips]{graphicx}

\makeatletter

\usepackage{babel}
\makeatother
\begin{document}

\title{Abundance of regular orbits and out-of-equilibrium phase transitions 
in the thermodynamic limit for long-range systems}


\author{R. Bachelard$^{1}$,
C. Chandre$^{1}$,
D. Fanelli$^{2}$,
X. Leoncini$^{1}$, 
S. Ruffo$^{2}$}

\affiliation{1. Centre de Physique Th\'{e}orique
, CNRS - Aix-Marseille Université, Luminy, Case 907, F-13288 Marseille
cedex 9, France \\
2.  Dipartimento di Energetica \char`\"{}Sergio Stecco\char`\"{}, 
Universit\'a
di Firenze, via s. Marta 3, 50139 Firenze, Italia and Centro 
interdipartimentale
per lo Studio delle Dinamiche Complesse (CSDC) and INFN}

\begin{abstract}
We investigate the dynamics of many-body long-range interacting systems, taking the Hamiltonian Mean Field model 
as a case study. We show that regular trajectories, associated with invariant tori of the 
single-particle dynamics, prevail. The presence of such tori provides a dynamical interpretation of the emergence 
of long-lasting out-of-equilibrium regimes observed generically in long-range systems. This is alternative to
a previous statistical mechanics approach to such phenomena which was based on a maximum entropy principle. 
Previously detected out-of-equilibrium phase transitions are also reinterpreted within this framework.
\end{abstract}


\maketitle

The vast majority of phenomena observed in nature results from complex interactions present in large assemblies of elementary constituents.
A widespread observation is the emergence of regular trajectories despite the complexity of the underlying network of couplings. A successful approach to describe the collective behaviour of  large  assemblies of particles is 
traditionally provided by statistical mechanics. The theoretical foundation of equilibrium 
statistical mechanics relies upon the hypothesis of ergodicity, i.e. the agreement
of time with ensemble averages. Arguing for an effective degree of global mixing of dynamical trajectories in phase space
implies ergodicity and thus the validity of statistical mechanics \cite{gallavotti}. Thermodynamic behaviour is obtained 
in the limit where the number of degrees of freedom goes to infinity, which offers innumerable pathways to chaos. 
Indeed, in this limit regular regions (invariant tori) 
do not possess enough dimensions in phase space to prevent trajectories from spreading,  while the largest fraction of the phase space is occupied
by chaotic motion, hence sustaining mixing \cite{froeschle}.

However, the fact that any weak nonlinearity would imply ergodicity has been vigorously debated since the pioneering work 
of Fermi, Pasta and Ulam (FPU)~\cite{FPU} on the dynamics of oscillators interacting via {\it short-range couplings}. 
Contrary to expectations, the celebrated FPU chain exhibits a recurrent behavior on very long times, violating 
ergodicity. Nowadays, there is a growing evidence that, for generic initial conditions, the relaxation time to equilibrium
remains finite in the thermodynamic limit~\cite{Gallavotti}.
On the contrary, the question of relaxation to equilibrium is still open when {\it long-range forces} come into 
play~\cite{Leshouches,Assisi}.
Indeed, systems with long-range interactions have been shown to display an extremely slow relaxation to 
equilibrium. More specifically, out-of-equilibrium metastable regimes have been identified, where the 
system gets trapped before eventually attaining its asymptotic state \cite{latora, yoshi}.
The equilibration time increases with system size and formally diverges in the thermodynamic limit, leading
to a breaking of ergodicity. For gravitational systems the approach to equilibrium
has never been proven and seems problematic.  Galaxies could therefore 
represent the most spectacular example of such far-from-equilibrium processes \cite{chavanis}, but analogous phenomena have also 
been reported in fundamental problems of plasma physics~\cite{Escande}. 
These metastable states have been termed Quasi-Stationary States (QSS) in the literature and often represent the solely 
accessible experimental regimes (e.g., in the Free Electron Laser~\cite{FEL_PRE}). 
A relevant feature of long-range  systems is the self-consistent nature
of the interaction of a particle with its "local field", which itself results from the combined action of all 
the other particles or of an "external field" like for wave-particle systems ~\cite{Escande}. 
It is exactly this self-consistency which finally entails the widespread regularity of the motion.
Besides that, one observes the presence of classes of initial states that 
show different time evolutions. 

The development of a systematic theoretical treatment of the QSS, 
which would enable us to unravel the puzzle of 
their ubiquity, is still an open problem. Both the self-consistent nature of the interaction
and the strong dependence on the initial condition suggest that QSS could be related 
to the presence of some type of regular motion. The traditional approach to clarifying
the emergence of regular trajectories is based on the following result~: 
If the Hamiltonian system under scrutiny is close to integrable, Kolmogorov-Arnold-Moser~\cite{KAM} theory
proves that the phase space is filled with invariant tori on which the motion is quasi-periodic. In this framework, 
however, increasing
the number of particles enhances the contribution of chaotic
trajectories~\cite{froeschle}, in stringent contradiction with the observation that QSS
prevail in the large $N$ limit. Therefore the aforementioned scenario cannot be invoked to 
explain the presence of regular motion in systems with long range interactions.

A first purpose of this Letter is to put forward a different interpretative framework. We 
argue that tori can form in phase space also as a result of a self-consistent interaction
in the thermodynamic limit. As we shall demonstrate, while for a small number $N$ of degrees
of freedom the single particle motion of a paradigmatic system with long-range interaction 
is erratic, the trajectories become more and more regular as $N \to \infty$. These trajectories 
arise from a low-dimensional time dependent effective Hamiltonian.

For systems with long-range interactions,
the dependence on the initial condition can materialize
in the form of a true out-of-equilibrium phase transition: By varying some crucial
parameters of the initial state, one observes a convergence towards asymptotic states  (in the limit
$N \to \infty$) with different macroscopic properties (e.g., homogeneous/inhomogeneous)~\cite{Anto1}.
This phase transition has been interpreted by resorting to Lynden-Bell's theory of "violent relaxation"~\cite{LB},
which is briefly discussed in the following with reference to a specific model. As a second purpose of this Letter, we provide a microscopic, dynamical, interpretation of this transition in terms of a
sharp modification of the properties of the single particle orbits, corresponding to a change of the
effective Hamiltonian. Our ultimate aim is to suggest a unifying  picture that potentially
applies to systems where collective, organized, phenomena emerge from
the globally coupled sea of individual components. 

The Hamiltonian Mean Field (HMF) model~\cite{anto95} is widely referred to as the benchmark 
for long-range systems and analysed for pedagogical reasons.
The model, which describes the evolution of $N$ particles coupled through an equally 
strong, attractive, cosine interaction is specified by the following Hamiltonian:
\begin{figure}
\begin{centering}
\includegraphics[width=9cm]{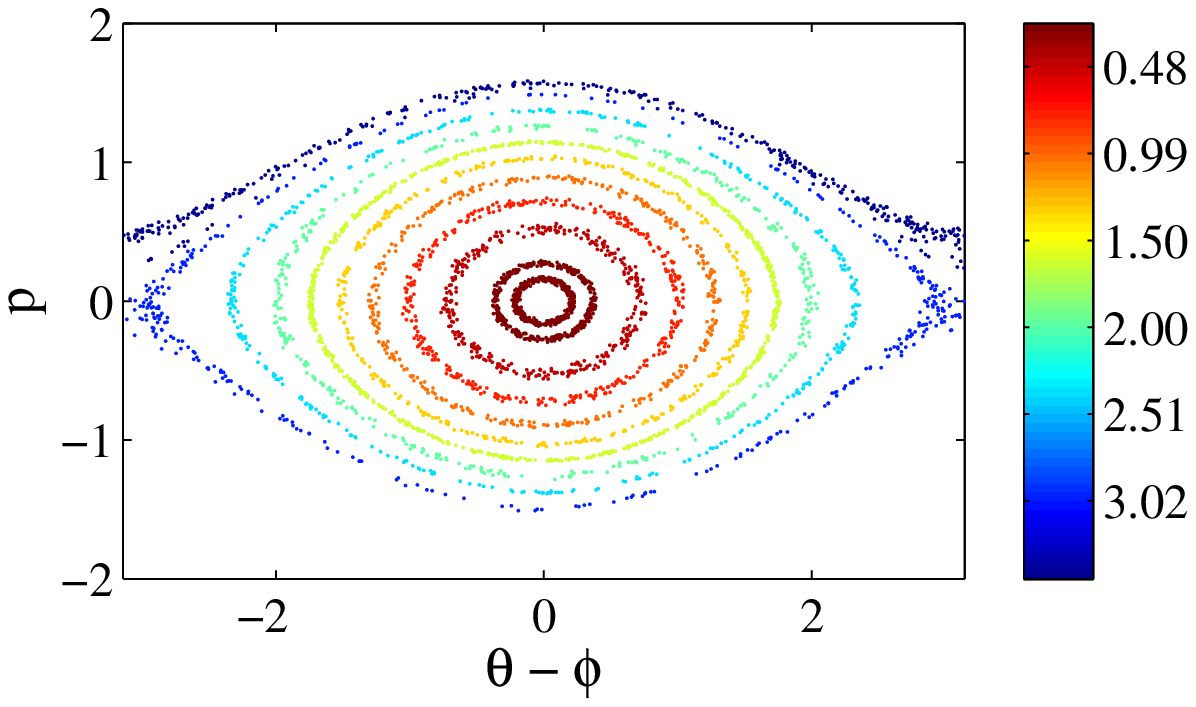}\\
\includegraphics[width=9cm]{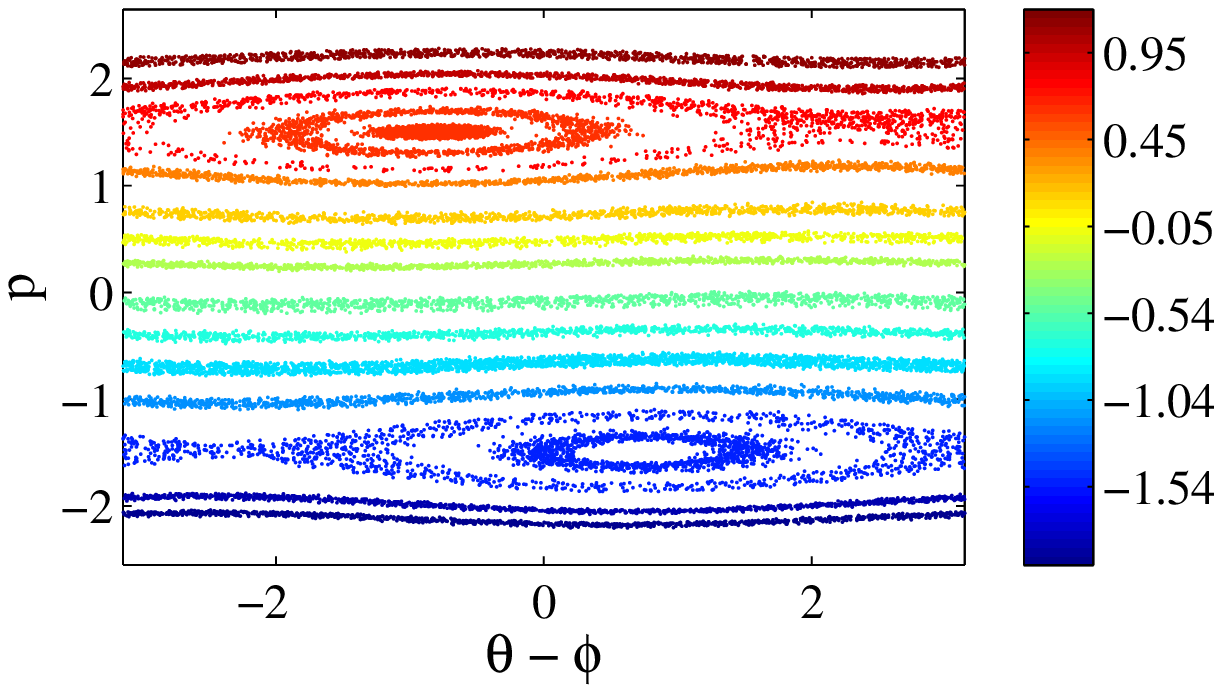}
\par\end{centering}
\caption{Poincar\'e sections of a few selected particles of one trajectory of Hamiltonian~(\ref{eq:Hamiltonian_HMF}) for $N=2\times 10^{5}$ in the 
QSS regime for two different water-bag initial conditions~: $(M_0,U)=(0.6,0.54)$ 
(upper panel) and $(M_0,U)=(0.6,0.88)$ (lower panel). 
The former returns a single cluster, which gives a non-zero magnetization QSS ($M_{QSS} \approx 0.5$),
while the latter shows two symmetric clusters, which produce a QSS with a small magnetization.
In the bicluster regime (lower panel) the presence of a large set of rotational tori implies a substantially lower
magnetization level, whereas the librational tori around the two clusters are responsible for the residual
magnetization. The color code corresponds to the values of the action variable associated with individual 
particles. 
\label{fig:Poincare_section}}
\end{figure}
\begin{equation}
H=\sum_{i=1}^{N}\left[\frac{p_{i}^{2}}{2}+\frac{1}{2N}\sum_{j=1}^{N}
\left(1-\cos\left(\theta_{i}-\theta_{j}\right)\right)\right],
\label{eq:Hamiltonian_HMF}
\end{equation}
where $\theta_i$ and $p_i$ label respectively the position of particle $i$ on the unit circle and its
corresponding momentum.
Note that Hamiltonian (\ref{eq:Hamiltonian_HMF}) can also be seen as a simplified version of the
gravitational~\cite{feix} or plasma~\cite{dawson} sheets model when considering only the first harmonic in the Fourier expansion of the potential. 
To characterize the behaviour of the system, it is convenient to introduce the ``magnetization''
$\mathbf{M}=\frac{1}{N}\left(\sum\cos\theta_{i},\:\sum\sin\theta_{i}\right)=M(\cos\phi,\:\sin\phi)$, which
quantifies the degree of spatial bunching of the particles (homogeneity vs. inhomogeneity).  
We here consider water-bag initial conditions, consisting of particles uniformly distributed in a  
rectangle $[-\theta_{0},\: \theta_{0}]\times[-p_{0},\: p_{0}]$ in the $(\theta,p)$-plane. These states bear a magnetization 
$M=M_{0}=\sin(\theta_{0})/\theta_{0}$, the associated energy per particle being $U=p_{0}^{2}/6+(1-M_{0}^{2})/2$. 
When performing numerical simulations, starting from the water-bag initial condition,
the system gets usually frozen in a QSS of the type discussed above \cite{latora,yoshi}. 

The individual particle $i$ obeys the following equations of motion
$
\dot{p_{i}} =  -M\:\sin\left(\theta_{i}-\phi\right)$ and $
\dot{\theta_{i}} = p_{i}$,
where $M$ and $\phi$ are functions of all the positions of the 
particles. Numerical observations suggest that, for large enough values of $N$, 
both the magnetization $M$ and the phase $\phi$ develop a specific oscillatory time dependence.
Hence, the single particle motion turns out to be governed by a time dependent one degree of freedom (often referred to as a one and a half degree of freedom)
effective Hamiltonian. 
This justifies the investigation of the phase space properties of the QSS using 
a technique inspired by that of Poincar\'e sections. 
In particular, we consider the time average $\bar{M}$ (after a 
transient) and record the positions and momenta of a few selected particles 
$(\theta_{i},\: p_{i})$ when $M(t)=\bar{M}$ and $dM/dt>0$ (since $M$
typically shows an oscillatory behavior). 
The resulting stroboscopic sections are displayed in Fig.~\ref{fig:Poincare_section}. 
Two different phase space structures are found depending on the choice of the
initial pair $(M_0,U)$, one with monocluster and the other with a bicluster.
The monocluster QSS displays a nonzero magnetization (inhomogeneous phase), while the bicluster QSS has a
small residual magnetization (homogeneous phase).

The monocluster QSS can be ideally mapped onto a collection of weakly interacting pendula. 
As revealed by our stroboscopic analysis, particles evolve on regular tracks, which are approximately 
one-dimensional, though they do manifest a degree of local diffusion ({\it thickness}). 
For the bicluster QSS, the Poincar\'e section shows a phase portrait which closely resembles the one obtained for a 
particle evolving in the potential of two contra-propagating waves. These latter
interact very weakly, as the associated propagation velocities appear rather different.

\begin{figure}
\begin{centering}
$\begin{array}{cc}
\includegraphics[width=4.0cm]{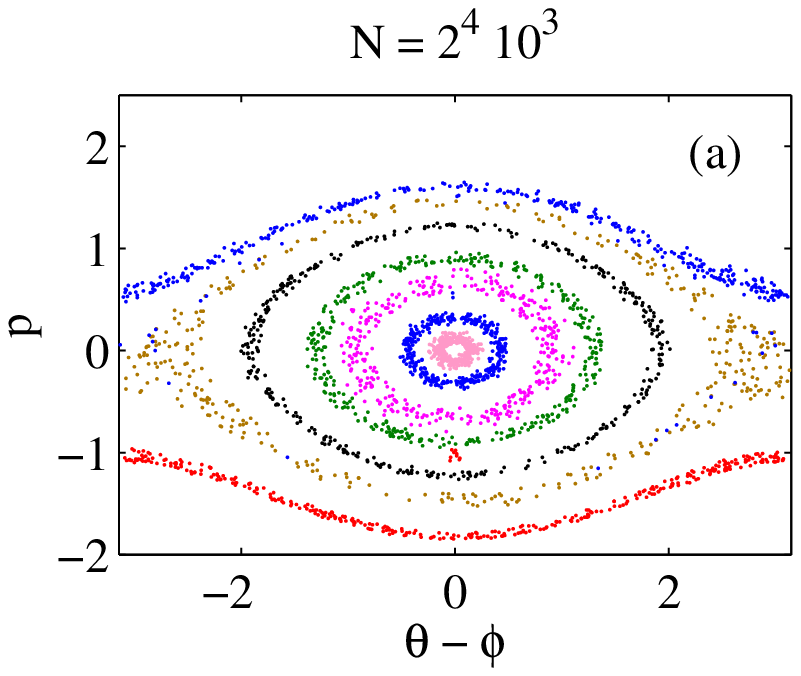}&\includegraphics[width=4.0cm]{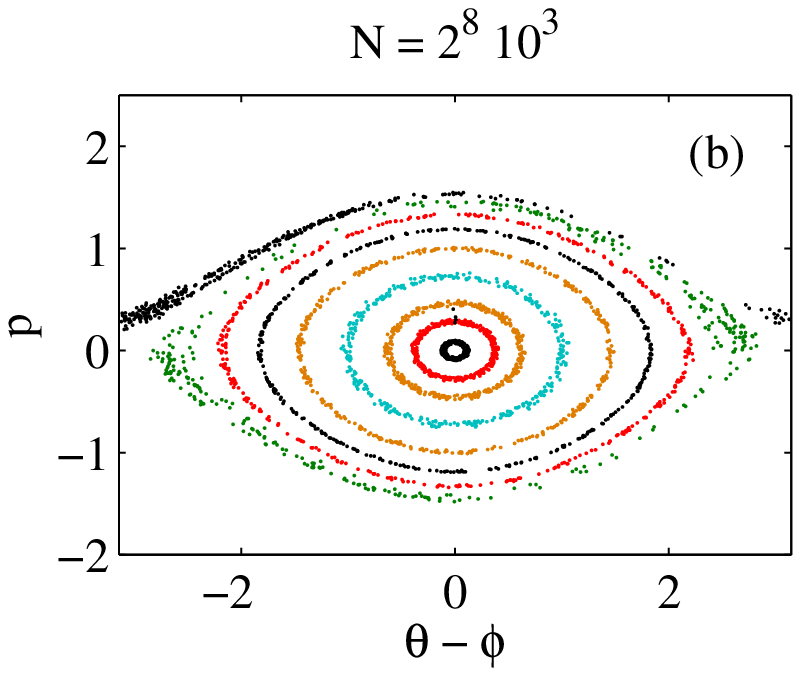}\\
\includegraphics[width=4.0cm]{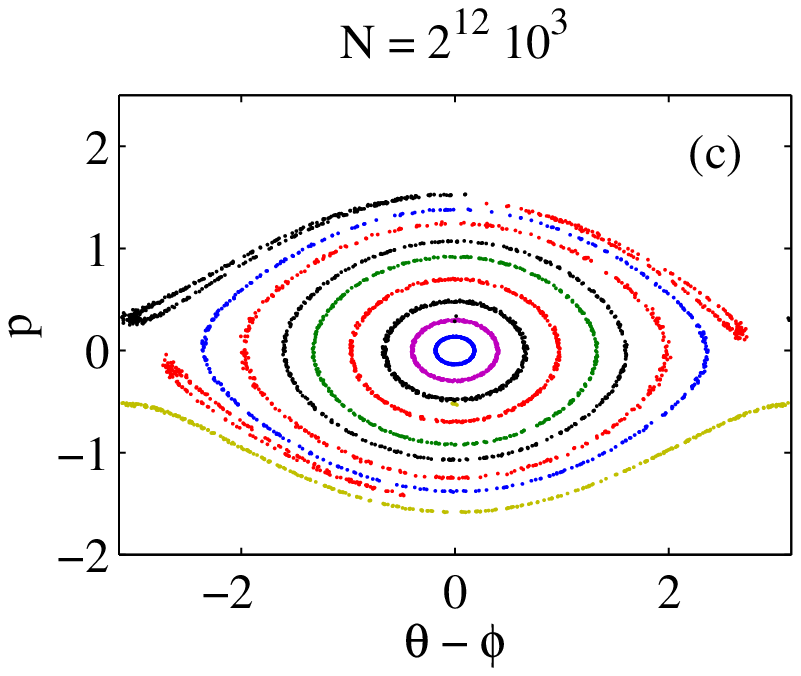}& 
\includegraphics[width=4.0cm]{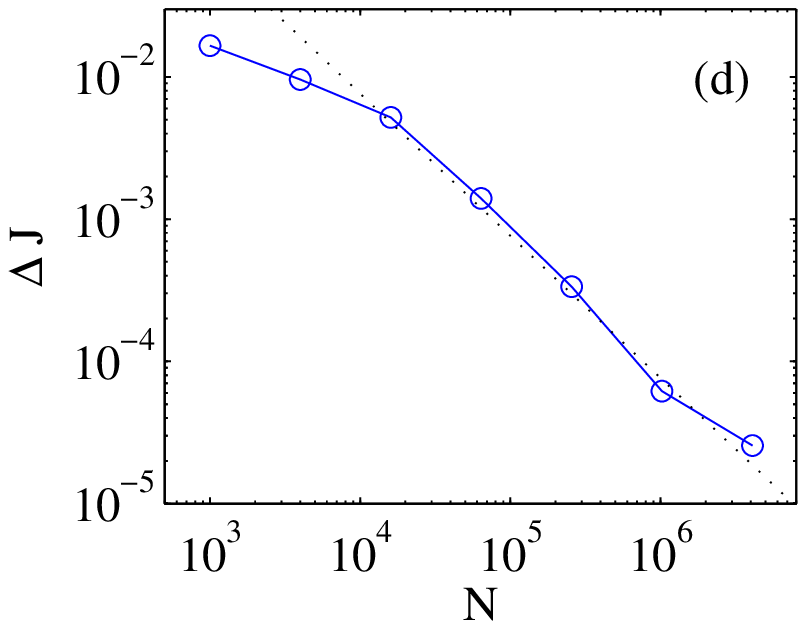}
\end{array}$
\par\end{centering}
\caption{Poincar\'e sections of a few selected particles of one trajectory of Hamiltonian 
(\ref{eq:Hamiltonian_HMF}), when the system size is varied (for $M_0=0.6$ 
and $U=0.54$). The {\it thickness} of the tori decreases as $N$ is increased (see text).
For large enough values of $N$, the magnetization $M$ is found numerically to approximately scale as 
$M(t)\approx \bar{M}+\delta M(t) \cos \omega t$, with $|\delta M|\ll \bar{M}$ and
$|\partial_t \delta M| \ll \omega |M|$.  
Ignoring the time dependence of $\delta M$ and using a reduced model 
of test particles in the external field $M(t)$, one obtains stroboscopic sections which are qualitatively and 
quantitatively similar to the ones reported in this figure, with the unique difference that the thickness
is zero \cite{Bachelard07}.
Considering a torus with action $J\approx 1.9$, we plot in panel d) its
variance $\Delta J$ computed over a time interval $\Delta t=300$ as a function of $N$.
The scaling $1/N$ (dotted line) looks accurate over a wide range of $N$ values. 
\label{fig:4sec}}
\end{figure}

In order to get a quantitative estimate of the {\it thickness} of the tori as a 
function of the total number of particles, we focus on the monocluster QSS.
Panels a), b) and c) in Fig.~\ref{fig:4sec} display the single particle phase
space for increasing values of $N$. A clear trend towards integrability is 
observed as quantified in panel d), where the {\it thickness} is plotted versus $N$.

Summing up, we have assessed that the single particle motion of a typical long-range interacting system 
becomes progressively more regular as the number of particles is increased. This
is at variance with what happens for systems with short-range interactions and provides 
a different interpretation of the abundance of regular motion in long-range 
dynamics.
Besides that, we have seen that the features of the single particle motion depend on the choice of
the initial condition. A natural question then arises: what is the link between the macroscopic
properties of the different QSS with the change observed in the single particle dynamics?
Anticipating the answer, we will see that this is related to a bifurcation
occurring in the effective Hamiltonian.

In the thermodynamic limit, the evolution of the single particle distribution function $f(\theta,p,t)$
is governed by the Vlasov equation~\cite{vlasovia}. This equation also describes 
the mean-field limit of wave-particle interacting systems~\cite{Escande}. 
It can be reasonably hypothesized that QSS are stationary stable solutions of 
the Vlasov equation~\cite{yoshi}. Following these lines, 
a maximum entropy principle, previously developed in the astrophysical context by Lynden-Bell~\cite{LB}, 
allowed one to predict \cite{Anto1,chava2}, for the HMF model, the occurrence of out-of-equilibrium phase transitions, 
separating distinct macroscopic regimes (magnetized/demagnetized) by varying selected control
parameters which represent the initial condition.

\begin{figure}
\begin{centering}
\includegraphics[width=9cm]{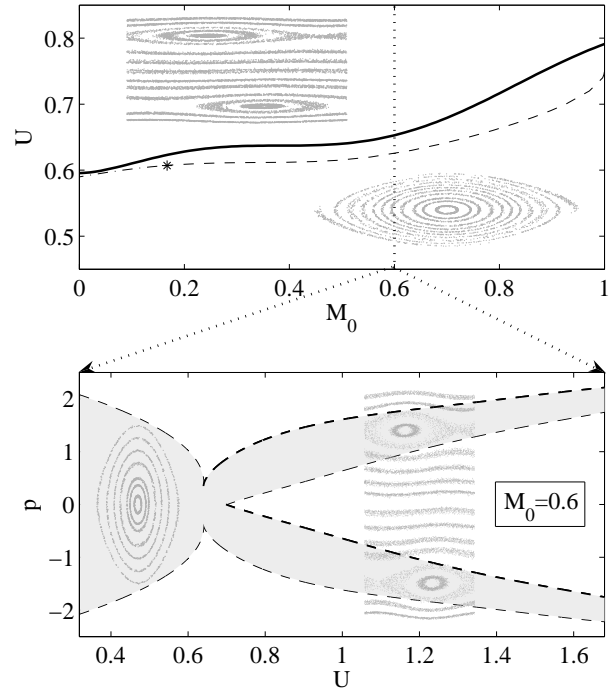}
\par\end{centering}
\caption{
Upper panel: Phase diagram in the control parameter plane $(M_{0},U)$ of the out--of--equilibrium phase
transition of the HMF model from a magnetized to a demagnetized phase. 
The solid curve pinpoints the position of the bifurcation from the monocluster to the bicluster QSS.
The dashed line stands for the theoretical prediction based on Lynden-Bell's violent relaxation theory (see text). 
The star refers to the tricritical point separating first from second order phase transitions. 
Lower panel: The bifurcation is monitored as function of $U$, for $M_0=0.6$. The grey zones highlight 
the width of the resonances.
\label{fig:transition}}
\end{figure}

The central idea of Lynden-Bell's approach consists in coarse-graining the microscopic one-particle distribution 
function $f(\theta,p,t)$ by introducing a local average in phase space. Starting from a water-bag initial 
profile, with a uniform distribution $f_0$, a fermionic-like entropy can be associated with the 
coarse grained profile $\bar{f}$, namely 
\begin{equation}
s[\bar{f}]=-\int \!\!{\mathrm d}p{\mathrm d}\theta \,\left[\frac{\bar{f}}{f_0} \ln \frac{\bar{f}}{f_0} 
+\left(1-\frac{\bar{f}}{f_0}\right)\ln \left(1-\frac{\bar{f}}{f_0}\right)\right]~.
\label{Lyndenbellentropy}
\end{equation}
The  corresponding statistical equilibrium, which applies to the relevant QSS regimes, is 
hence determined by maximizing such an entropy, while imposing the conservation of the Vlasov dynamical 
invariants, namely energy, momentum and norm of the distribution. The analysis reveals 
the existence of an out-of-equilibrium phase transition from a magnetized to a demagnetized phase
\cite{Anto1,chava2}.

We here reinterpret the transition in a purely dynamical framework, as a bifurcation
from a monocluster QSS to a bicluster QSS. Aiming at shedding light on this issue, we 
proceed as follows: For fixed
$M_{0}$ and $N$, we gradually increase the energy $U$ and compute
the Poincar\'e sections, as discussed above. We then analyze the recorded sections by identifying the number of 
resonances and measuring the associated width and position (both calculated in the $p$ direction).
Results for $M_{0}=0.6$ are displayed in the lower panel of Fig.~\ref{fig:transition}:
the shaded region, bounded by the dashed lines, quantifies the width of the resonances.
As anticipated, one can recognize the typical signature of a
bifurcation pattern. Repeating the above analysis for different values of the initial magnetization $M_0$,
allows us to draw a bifurcation line in the parameter space $(M_0, U)$. 
In the upper panel of Fig.~\ref{fig:transition} we report both this bifurcation 
(full) and the Lynden-Bell phase transition (dashed) lines \cite{Anto1}.
The two profiles resemble each other qualitatively, and even quantitatively for small $M_0$. 
The change of the bifurcation type from subcritical to supercritical signaled by the opening
of a gap in the resonance plot could be associated with the
change of the order of the phase transition from first to second.
These results strongly corroborate our claim that the out--of--equilibrium phase 
transition from magnetized to non magnetized QSS corresponds to a bifurcation in the
single particle dynamics.

The phenomena here discussed are also found (data not shown) in the context of Hamiltonian models deputed 
to describe the interaction between a beam of charged particles and a set of 
self-consistently evolving waves \cite{bonifacio}. A large number of relevant 
applications, facing the long-range nature of the couplings among constituents,  
fall within this rather wide category, ranging from plasma models,  
traveling wave tubes, and free electron lasers \cite{FELscience}. More concretely, the same 
tendency towards integrability is found in these latter settings, a fact which holds  
promise to eventually result in novel insight, potentially relevant for the functioning of the 
devices. 

In summary, investigating the dynamics of systems with long range interactions allowed us to come to an
interesting and general conclusion: a universal trend to organization/integrability of the single particle dynamics is 
found as the system size (number of particles) is increased. Phase space structures are 
identified and interpreted as invariant tori of a time dependent one degree of freedom Hamiltonian.
Based on this analysis, the out-of-equilibrium statistical phase transitions 
is here understood from the viewpoint of microscopic dynamics, as a bifurcation 
occurring in the effective Hamiltonian which describes the single particle dynamics.

\end{document}